\documentclass[12pt]{article}

\title{Recursion for Masur-Veech volumes of moduli spaces of quadratic differentials}
\author{Maxim Kazarian\thanks{
National Research University Higher School of Economics,
Skolkovo Institute of Science and Technology, e-mail: kazarian@mccme.ru}}

\date{}

\usepackage{amssymb,amsmath}
\usepackage{theorem}

\newtheorem{theorem}{Theorem}

{\theorembodyfont{\rmfamily}    
\newtheorem{remark}[theorem]{Remark}

}

\def\mdskip{\ifnum\medskipamount>\lastskip
      \vskip-\lastskip\vskip\medskipamount\fi}
\def\QED{\ifmmode\eqno\square\else
{\parfillskip0pt\hfil$\square$\par}\mdskip\fi}

\def\ocM{{\overline{{\cal M}}}}
\let\l\lambda
\let\t\tau
\let\d\delta
\let\e\epsilon

\let\D\partial
\def\pd#1#2{\frac{\partial#1}{\partial#2}}

\begin{document}

\maketitle

\begin{abstract}
We derive a quadratic recursion relation for the linear Hodge integrals of the form $\langle\t_2^n\lambda_k\rangle$. These numbers are used in a formula for Masur-Veech volumes of moduli spaces of quadratic differentials discovered by Chen, M\"oller, and  Sauvaget in~\cite{CMS}. Therefore, our recursion provides an efficient way of computing these volumes.
\end{abstract}

\section{The recursion}

We use Witten's notation for linear Hodge integrals over the moduli space $\ocM_{g,n}$ of stable genus~$g$ curves with $n$ marked points,
$$\langle\t_{i_1}\dots\t_{i_n} \l_k \rangle=\int_{\ocM_{g,n}}\l_k\psi_1^{i_1}\dots\psi_n^{i_n}$$
where $\psi_i$ and $\l_k$ are the corresponding standard characteristic classes and~$g$ is recovered from dimension count, $3g-3+n=k+\sum i_j$.
In this note we derive a quadratic recurrent relation for the numbers of the form $\langle\t_2^m\l_k\rangle$. The numbers of this sort are involved in a formula for Masur-Veech volume of the principal stratum of moduli space of quadratic meromorphic differentials derived in~\cite{CMS}. Namely, with the above notation this formula reads
\begin{align*}
V_{g,n}&=\text{vol}(Q_{g,4 g-4+2 n}(1^{4 g-4+n},-1^n))\\
&=2^{2 g+1} \pi ^{6 g-6+2 n}\frac{(4 g-4+n)!}{(6 g-7+2 n)!}
\sum _{k=0}^g \frac{ \langle \tau _0^n  \tau _2^{3 g-3+n-k}\l_{k}\rangle}{(3 g-3+n-k)!}\\
&=2^{2 g+1} \pi ^{6 g-6+2 n}\frac{(4 g-4+n)!}{(6 g-7+2 n)!}\frac{(5 g-7+2 n-k)!!}{(5 g-7-k)!!}
\sum _{k=0}^g \frac{ \langle \tau_2^{3 g-3-k}\l _{k} \rangle}{(3 g-3-k)!}
\end{align*}
The second equality holds for $g\ge2$ and follows from the first one by the string and dilaton relations allowing one to eliminate $\t_0$ and $\t_1$ from correlators:
$$\langle  \tau _0^{m+1} \tau _2^{n+1}\l_k\rangle
=(n+1)\langle \tau _1  \tau _0^m \tau _2^n\l _k\rangle
=(n+1)\,(2 g-2+m+n) \langle  \tau _0^m \tau _2^n\l _k\rangle,
$$
where $n-m=3 g-3-k$. There are some corrections in the string and dilaton relations in the cases $g=0$ and $g=1$ due to absence of the moduli spaces $\ocM_{0,\leq 2}$ and $\ocM_{1,0}$. Fortunately, the corresponding intersection numbers for $g\le1$ are easy to compute explicitly: by the string and dilaton they are reduced to just $\langle\t_0^3\rangle=1$ and $\langle\t_1\rangle=\langle\t_0\l_1\rangle=\frac{1}{24}$ and we recover the known values (cf.~\cite{AEZ} for the case~$g=0$ and~\cite{CMS} for the case~$g=1$)
$$V_{0,n}=\frac{\pi ^{2 n-6}}{2^{n-5}},
\qquad
V_{1,n}=\frac{\pi ^{2 n}\, n!}{3 (2 n-1)!} \bigl((2 n-3)!!+(2 n-2)!!\bigr).$$

The presented below recursion for the numbers $\langle\t_2^m\l_k\rangle$ looks somewhat nicer with the following normalization. Denote
$$c_{g,k}=\frac{\langle \tau _0^2  \tau _2^{3 g-1-k}\l _k\rangle }{(3 g-1-k)!}
=(5 g-3-k)(5g-5-k)\frac{\langle \l _k \tau _2^{3 g-3-k}\rangle }{(3 g-3-k)!}.$$
These numbers are nonzero only if $g\ge 1$ and $0\le k\le g$.

\begin{theorem}\label{th1} We have $c_{1,0}=\frac1{12}$ and for all other pairs $(g,k)$ the following recursion relation holds
$$c_{g,k}=\tfrac{g+1-k}{5 g-2-k} c_{g,k-1}+\tfrac {(5 g-6-k) (5 g-4-k)}{12} c_{g-1,k}+
\frac{1}{2}\sum_{\substack{g_1+g_2=g\\k_1+k_2=k}} c_{g_1,k_1}c_{g_2,k_2}.$$
\end{theorem}
The computer allows one to find the constants $c_{g,k}$ by this recursion and thus, the volume
$$V_{g,0}=\frac{2^{2 g+1} \pi ^{6 g-6} (4 g-4)!}{(6 g-7)!}\sum_{k=0}^g\frac{c_{g,k}}{(5 g-3-k)(5g-5-k)}$$ up to, say, $g=100$ in just few seconds, which helps to study experimentally the large genus asymptotics of volumes. For example, one can observe numerically that the ratio between the exact value of $V_{g,0}$ and its conjectural asymptotic value
$$V_{g,0}\approx\frac{4}{\pi}\left(\frac{8}{3}\right)^{4g-4}$$
for $g=100$ is equal  to $0.9993$ supporting a conjecture from~\cite{DGZZ}. According to~\cite{CMS}, the work in progress~\cite{YZZ} will contain a refined version of this conjecture.

I thank Anton Zorich for pointing me the problem and for his patient explanation of the background from which the problem originated. The research of the author is supported by the Russian Science Foundation (project 16-11-10316).

\section{Proof of Theorem~1}
To simplify the arguments, consider first the case $k=0$, that is, we compute first the numbers
$$c_g=c_{g,0}=\frac{\langle\t_0^2\t_2^{3g-1}\rangle}{(3g-1)!}=(5g-3)(5g-5)\frac{\langle\t_2^{3g-3}\rangle}{(3g-3)!}.$$
(The rightmost equality holds for $g>1$ only while for $g=1$ we have $c_{1}=c_{1,0}=\frac1{12}$).
Consider the Kontsevich-Witten potential which is the generating function for $\t$-correlators
$$F(\hbar;t_0,t_1,\dots)=\sum_{g,n}\hbar^g\sum_{i_1+\dots+i_n=3g-3+n}\langle\t_{i_1}\dots\t_{i_n}\rangle\frac{t_{i_1}\dots t_{i_n}}{n!}.$$
By Kontsevich-Witten theorem, it satisfies the equations of KdV hierarchy, in particular, the KdV equation itself (see~\cite{W}, \cite{Ko},~\cite{IZ}),
 $$\pd{^2F}{t_0\D t_1}=\frac12\Bigl(\pd{^2F}{t_0^2}\Bigr)^2+\frac{\hbar}{12}\pd{^4F}{t_0^4}$$
Let us restrict this equation to the point $(t_0,t_1,t_2,\dots)=(0,0,1,0,\dots)$. Namely, set
$$f_{k_1\dots k_n}=\pd{^nF}{t_{k_1}\dots\D t_{k_n}}\bigm|_{t_i=\d_{i,2}}.$$
Then we have
\begin{equation}\label{KdV}
 f_{01}=\frac12f_{00}^2+\frac{\hbar}{12}f_{0000}.
\end{equation}

 Now we observe that the coefficients of the series involved in this equation contain only intersection numbers of the form $\langle\t_0^i\t_1^j\t_2^n\rangle$ with small~$i$ and~$j$, and hence, they can be expressed in terms of $c_g$. Explicitly, applying repeatedly the string and dilaton relations, we get
 \begin{align*}
 f_{00}&=\sum_{g\ge1}\frac{\langle\t_0^2\t_2^{3g-1}\rangle}{(3g-1)!}\hbar^g=\sum_{g\ge1}c_g\hbar^g,\\
 f_{01}&=\sum_{g\ge1}\frac{\langle\t_0\t_1\t_2^{3g-2}\rangle}{(3g-2)!}\hbar^g=\sum_{g\ge1}\frac{\langle\t_0^2\t_2^{3g-1}\rangle}{(3g-1)!}\hbar^g=f_{0,0},\\
 \hbar f_{0000}&=\sum_{g\ge1}\frac{\langle\t_0^4\t_2^{3g-2}\rangle}{(3g-2)!}\hbar^g
 =\sum_{g\ge2}(5g-4)(5g-6)\frac{\langle\t_0^2\t_2^{3g-4}\rangle}{(3g-4)!}\hbar^g+\hbar\\
 &=\sum_{g\ge2}g(5g-4)(5g-6)c_{g-1}\hbar^g+\hbar=(5G-4)(5G-6)(\hbar f_{0,0})+\hbar,
 \end{align*}
where $G=\hbar\frac{\D}{\D\hbar}$. Substituting to the KdV equation~\eqref{KdV} we get an equation on the generating series $f_{00}$ for the numbers $c_g$,
$$f_{00}=\frac{(5G-4)(5G-6)}{12}(\hbar f_{00})+\frac12f_{00}^2+\frac\hbar{12}$$
which is equivalent to the recursion:
$$c_g=\frac{(5g-4)(5g-6)}{12}c_{g-1}+\frac12\sum_{g_1+g_2=g}c_{g_1}c_{g_2}.$$
(See~\cite{IZ}, \cite{MZ},~\cite{Z} where this recursion is derived by essentially the same arguments).

\bigskip
In the general case, we consider similarly the generating function for linear Hodge integrals,
$$H(\hbar,s;t_0,t_1,\dots)=\sum_{g,n}\hbar^g\sum_{i_1+\dots+i_n+k=3g-3+n}\langle\t_{i_1}\dots\t_{i_n}\l_k\rangle
\; s^k\;\frac{t_{i_1}\dots t_{i_n}}{n!}.$$
Set $s=-\e^2$ and consider a sequence of linear functions $T_k(p_1,p_2,\dots)$ defined recursively by
$$T_0=p_1,\qquad T_{k+1}=\sum_m m\,(p_{m+2}+\e\,p_{m+1})\frac{\D T_k}{\D p_m}.$$
We have
\begin{align*}
T_1=&\,\e\, p_2  +p_3,\\
T_2=&2\,\e^2 p_3+5 \,\e\, p_4 +3 \,p_5,\\
T_3=&6 \,\e^3p_4 + 26 \,\e^2p_5+35 \,\e\, p_6 +15\, p_7,\\
T_4=&24 \,\e^4p_5 +154 \,\e^3p_6 + 340 \,\e^2p_7+315 \,\e\, p_8 +105\, p_9,
\end{align*}
and so on. The substitution $t_k=T_k(p)$ is a polynomial triangular invertible (for $\e\ne0$) change of variables.

\begin{theorem}[\cite{K}]\label{KP}
The function $H$ written in $p$~coordinates satisfies equations of KP hierarchy (for any value of $\e$).
\end{theorem}

\begin{remark} We used this fact in~\cite{K} to provide one of the shortest proofs of Kontsevich-Witten theorem. Indeed, setting $\e=0$ we get $T_k|_{\e=0}=(2 k-1)!! p_{2 k+1}$. Therefore, $F=H\bigm|_{\e=0}$ is a solution of KP hierarchy. Moreover, this solution is independent of even times $p_{2k}$, that is, it is a solution of the KdV hierarchy.

Another integrable hierarchy of partial differential equations for~$H$, namely, the Intermediate Long Wave hierarchy is discovered in~\cite{B}. Its equations are slightly more complicated but can also be used for derivation of recursion for $c_{g,k}$.
\end{remark}

The first equation of the KP hierarchy is the KP equation itself,
$$\frac{\D ^2H}{\D p_1\, \D p_3}=\frac{\D ^2H}{\D p_2^2}
+\frac{\hbar}{12} \frac{\D ^4H}{\D p_1^4}
+\frac{1}{2} \Bigl(\frac{\D ^2H}{\D p_1^2}\Bigr)^2.$$
By the change of coordinates described in Theorem~\ref{KP} we have
$$
\frac{\D H}{\D p_1}=\frac{\D H}{\D t_0},\qquad
\frac{\D H}{\D p_2}=-\epsilon  \frac{\D H}{\D t_1},\qquad
\frac{\D H}{\D p_3}=\frac{\D H}{\D t_1}-2 \epsilon ^2 \frac{\D H}{\D t_2}.
$$
Therefore, the KP equation in $t$-coordinates takes the following form (recall that $s=-\e^2$)
$$\frac{\D ^2H}{\D t_0\, \D t_1}=
2s\frac{\D ^2H}{\D t_0\D t_2}
-s\frac{\D ^2H}{\D t_1^2}
+\frac{\hbar}{12}   \frac{\D ^4H}{\D t_0^4}
+\frac{1}{2} \Bigl(\frac{\D ^2H}{\D t_0^2}\Bigr).$$
Let us restrict the KP equation to $t=(0,0,1,0,\dots)$. Setting
$$h_{k_1\dots k_n}=\pd{^nH}{t_{k_1}\dots\D t_{k_n}}\bigm|_{t_i=\d_{i,2}}$$
we get
\begin{equation}\label{KPeq}
h_{01}=2\, s\, h_{02}-s\, h_{11}+\frac{\hbar}{12}  h_{0000}+\frac12 h_{00}^2.
\end{equation}
The coefficients of all involved series include only intersection numbers of the form $\langle\t_0^i\t_1^j\t_2^n\l_k\rangle$, with small~$i$ and~$j$ and, hence, they can be expressed in terms of $c_{g,k}$. Applying repeatedly the string and dilaton relations, we get
 \begin{align*}
h_{00}&=\sum_{g,k}\frac{\langle\t_0^2\t_2^{3g-1-k}\l_k\rangle}{(3g-1-k)!}\hbar^gs^k=\sum_{g,k}c_{g,k}\hbar^gs^k,\\
h_{01}&=\sum_{g,k}\frac{\langle\t_0\t_1\t_2^{3g-2-k}\l_k\rangle}{(3g-2-k)!}\hbar^gs^k=
\sum_{g,k}\frac{\langle\t_0^2\t_2^{3g-1-k}\l_k\rangle}{(3g-1-k)!}\hbar^gs^k=h_{0,0},\\
s\,h_{0,2}&=\sum _{g,k} \frac{\langle \tau _0  \tau _2^{3 g-1-k}\l _{k-1}\rangle }{(3 g-2-k)!}\hbar ^g s^k
=\sum _{g,k} (3 g-1-k)\frac{ c_{g,k-1}}{5 g-2-k}\hbar ^g s^k,\\
s\,h_{1,1}&=\sum _{g,k} \frac{\langle \tau _1^2  \tau _2^{3 g-2-k}\l _{k-1}\rangle }{(3 g-2-k)!}\hbar ^g s^k
=\sum _{g,k} (5 g-3-k)  \frac{c_{g,k-1}}{5 g-2-k}\hbar ^g s^k,\\
s\,h_0&=\sum _{g,k} \frac{ \langle \tau _0  \tau _2^{3 g-1-k}\l _{k-1}\rangle }{(3 g-1-k)!}\hbar ^g s^k
=\sum _{g,k} \frac{c_{g,k-1}}{5 g-2-k}\hbar ^g s^k ,\\
\hbar\,h_{0,0,0,0}&=\sum _{g,k} \frac{ \langle \tau _0^4 \tau _2^{3 g-2-k}\l _k \rangle }{(3 g-2-k)!}\hbar ^g s^k
=\sum _{g,k} (5 g-4-k) (5 g-6-k)  c_{g-1,k}\hbar ^g s^k+\hbar.
 \end{align*}
Introduce commuting operators $G=\hbar\frac\D{\D \hbar}$,  $K=s\frac\D{\D s}$. Any operator of the form $P(G,K)$, where $P$ is a polynomial, acts on a series in $\hbar$ and $s$ multiplying any monomial of the form $\hbar^gs^k$ by $P(g,k)$. With these notations, the KP equation~\eqref{KPeq} takes the form of differential equation on the generating function $h_{0,0}$ for the coefficients $c_{g,k}$:
\begin{align*}
h_{0,0}=(G-K+1)(s\,h_0&)+\frac{(5 G-K-4) (5 G-K-6)}{12} (\hbar\,  h_{0,0})+\frac12h_{0,0}^2+\frac{\hbar }{12},\\[1ex]
s\,h_0&=(5 G-K-2)^{-1}(s\,h_{0,0}).
\end{align*}
Taking the coefficient of $\hbar^gs^k$ we get exactly the relation of Theorem~\ref{th1}.

\end{document}